\documentclass[aps,preprint,superscriptaddress,onecolumn,amsmath,floatfix,citeautoscript]{revtex4}
\usepackage{graphicx}
\usepackage{float}
\usepackage{dcolumn}
\usepackage{color}
\usepackage{latexsym,bm}
\usepackage[normalem]{ulem}
\usepackage{multirow}
\usepackage{appendix}
\usepackage{epstopdf}
\usepackage{booktabs}
\usepackage[colorlinks,linkcolor=blue,citecolor=blue,urlcolor=blue]{hyperref}

\begin{document}

\hyphenpenalty=5000
\tolerance=1000

\title{Nonlocal Model for Electron Heat Flux and Self-generated Magnetic Field}

\date{\today}

\author{Xinyu Zhu}
\affiliation{Center for Applied Physics and Technology, HEDPS and State Key Laboratory of Nuclear Physics and Technology, School of Physics, Peking University, Beijing 100871, People’s Republic of China}
\author{Wenqiang Yuan}
\affiliation{Center for Applied Physics and Technology, HEDPS and State Key Laboratory of Nuclear Physics and Technology, School of Physics, Peking University, Beijing 100871, People’s Republic of China}
\author{Yusen Wang}
\affiliation{Center for Applied Physics and Technology, HEDPS and State Key Laboratory of Nuclear Physics and Technology, School of Physics, Peking University, Beijing 100871, People’s Republic of China}
\author{Zhipeng Zhang}
\affiliation{Center for Applied Physics and Technology, HEDPS and State Key Laboratory of Nuclear Physics and Technology, School of Physics, Peking University, Beijing 100871, People’s Republic of China}
\author{Xianxu Jin}
\affiliation{Center for Applied Physics and Technology, HEDPS and State Key Laboratory of Nuclear Physics and Technology, School of Physics, Peking University, Beijing 100871, People’s Republic of China}
\author{Zhonghai Zhao}
\affiliation{Center for Applied Physics and Technology, HEDPS and State Key Laboratory of Nuclear Physics and Technology, School of Physics, Peking University, Beijing 100871, People’s Republic of China}
\author{Bin Qiao}
\affiliation{Center for Applied Physics and Technology, HEDPS and State Key Laboratory of Nuclear Physics and Technology, School of Physics, Peking University, Beijing 100871, People’s Republic of China}

\date{\today}

\begin{abstract}
Coupling of electron heat conduction and magnetic field takes significant effects in inertial confinement fusion (ICF). As the nonlocal models for electron heat conduction have been developed for modeling kinetic effects on heat flux in hydrodynamic scale, modeling kinetic effects on magnetic field are still restricted to flux limiters instead of nonlocal corrections. We propose a new nonlocal model which can recover the kinetic effects for heat conduction and magnetic field in hydrodynamic scale simultaneously. We clarify the necessity of self-consistently considering the electric field corrections in nonlocal models to get reasonable physical quantities. Using the new nonlocal model, the nonlocal corrections of transport coefficients in magnetized plasma and the magnetic field generation without density gradients are systematically studied. We find nonlocal effects significantly change the magnetic field distribution in laser ablation, which potentially influences the hydrodynamic instabilities in ICF.
\end{abstract}
\maketitle

\section{Introduction}
Coupling of magnetic field and electron heat conduction is of essentially importance in inertial confinement fusion (ICF)\cite{14PRL-Lancia,17PoP-Farmer,22PoP-Sadler}. Electron heat flux advects the magnetic field\cite{08PRL-Ridgers,23PRL-Arran} while magnetic field suppresses the electron heat flux and deflects the flux to the transverse direction of temperature gradient (Righi-Leduc heat flux)\cite{10PRL-Bissell,22MRE-Zhao}, thus influencing the plasma parameters\cite{17PRL-Walsh,21PRL-Matsuo}. This takes significantly effects in ICF including enhancing the confinement\cite{17PRL-Walsh,24PRR-Grandvaux}, adapting the symmetry of implosion\cite{22PRL-Bose}, and influencing the hydrodynamic instabilities\cite{12PRL-Manuel,21PRL-Matsuo,22PoP-Zhang,24JFM-Zhang}. Physically, the magnetic field evolution and heat conduction are coupled processes according to the kinetic principle\cite{84JPDA-Epperlein,24PoP-Lopez} and can be directly studied through kinetic simulations\cite{86PoF-Epperlein,21PRL-Sadler}, although unrealistic for large scale. As the temperature gradient is not too large, the classical transport coefficients for heat flux and magnetic field can be obtained together through the kinetic theory\cite{86PoF-Epperlein,20PoP-Walsh,21PRL-Sadler} and used in radiation hydrodynamic (RHD) simulations\cite{20PoP-Walsh,21PRL-Sadler}.

The classical transport theory becomes invalid with large temperature gradient and needs correction\cite{PICF-Atzeni}. For the heat conduction, it is well known that flux limit effects exist and flux limiter is used traditionally\cite{PICF-Atzeni}. In recent studies, nonlocal effects including peak flux suppression and preheating\cite{83PRL-Luciani,85PRL-Luciani}, which significantly influence ICF\cite{22MRE-Li,24PRR-Yuan} and can't be simply explained by flux limiter\cite{22PRL-Midler,24PPCF-Yuan}, have been confirmed in experiments\cite{07PRL-Froula,07PRL-Schurtz,22PRL-Midler}. To overcome the restriction of direct kinetic simulation and calculating the effects appropriately in RHD simulations, multiple nonlocal models\cite{00PoP-Schurtz,06PoP-Nicolai,15PoP-DelSorbp,16PoP-DelSorbo,08PoP-Wallace,18PoP-Manheimer,18PoP-Manheimer-2,06NJP-Frolov} based on kinetic theory are proposed to modeling the effects, which significantly improves the accuracy of heat conduction in RHD simulation\cite{24PPCF-Yuan,24PRR-Yuan}. On the other hand, nonlocal effects of magnetic field evolution are also observed in kinetic simulation and experiments recently, including the suppression of self-generated magnetic field\cite{20PRL-Sherlock,22PoP-Campbell,23PoP-Davies} and advection of magnetic field\cite{18PPCF-Brodrick,24PoP-Walsh}, and the non-classical mechanism of magnetic field generation\cite{02PRL-Kingham,04JCP-Kingham}. And some flux limiters are proposed to attempt to recover the effects\cite{20PRL-Sherlock,23PoP-Davies,24PoP-Walsh}.

While multiple nonlocal models have been developed to modelling the nonlocal effects for heat conduction in RHD simulations, appropriate nonlocal model considering magnetic field correction together has not been developed and the modelling of non-classical corrections for magnetic field is mostly restricted to the flux limiter in RHD simulations\cite{18PPCF-Brodrick,20PRL-Sherlock,23PoP-Davies,24PoP-Walsh}. Simply extending existing nonlocal model for heat conduction to magnetic field evolution usually gives unreasonable results as self-consistent consideration the correction of electric field doesn't receive attention. Direct kinetic simulations can get exact nonlocal corrections but are too expensive to extend to the hydrodynamic scale and are usually restricted to small temperature perturbation for the correction of classical transport coefficients\cite{95PRL-Bychenkov,03PoP-Bychenkov,04PoP-Doumaz}. 

In this paper, we propose a nonlocal model in magnetized plasma, which can recover the kinetic effects for heat conduction and magnetic field evolution simultaneously by self-consistently considering the correction of electric field. The paper is organized as follows. In Section 1, we will review the classical transport theory briefly. In Section 2, we will establish the nonlocal model and using a classical test to show the necessity of considering the electric field correction self-consistently. Finally, we will study the magnetic field generation without density gradients and study the laser ablation through the implementation of the model in RHD code FLASH.

\section{Classical transport theory in magnetized plasma}

Firstly we will review the classical transport theory, especially the exact expression of the Maxwellian anisotropic electron distribution function (EDF) $\textbf{f}_1^M$, which is basic for our non-local theory.

The basic equation for transport theory in magnetized plasma is the Fokker-Planck (FP) equation\cite{PICF-Atzeni,06PoP-Nicolai}
\begin{equation}
\frac{\partial f}{\partial t}+\textbf{v}\cdot\nabla f-\frac{e}{m}(\textbf{E}+\textbf{v}\times \textbf{B})\cdot\nabla_v f=C_{ee}+C_{ei}.
\end{equation}
$f$ is the EDF in phase space, and $v$ is the electron velocity. $e, m$ are electron charge and electron mass. $E, B$ are electric field and magnetic field. $C_{ee}, C_{ei}$ are electron-electron and electron-ion collision operators respectively. Using the first-order expansion of EDF in phase space\cite{PICF-Atzeni,06PoP-Nicolai}
\begin{equation}
f(\textbf{r},\textbf{v},t) = f_0(\textbf{r},v,t)+\frac{\textbf{v}}{v}\cdot \textbf{f}_1 (\textbf{r},v,t),
\end{equation}
here $f_0$ is the isotropic part of the EDF, and $f_1$ is the first-order anisotropic part of the EDF, and considering a steady state regime, we can simplify the original FP equation to the form\cite{06PoP-Nicolai}

\begin{subequations}

\begin{equation}
\label{3a}
    \frac{v}{3}\nabla\cdot \textbf{f}_1-\frac{e\textbf{E}}{3mv^2}\cdot \nabla_v(v^2 \textbf{f}_1)=C_{ee}^{0}(f_0),
\end{equation}
\begin{equation}
\label{3b}
v\nabla f_0 -\frac{e\textbf{E}}{m}\frac{\partial f_0}{\partial v}-\frac{e\textbf{B}}{m}\times\textbf{f}_1=C_{ei}^1(\textbf{f}_1)+C_{ee}^1(\textbf{f}_1).
\end{equation}
\end{subequations}

\begin{figure}[htbp]
\centering
\includegraphics[width=1.0\textwidth]{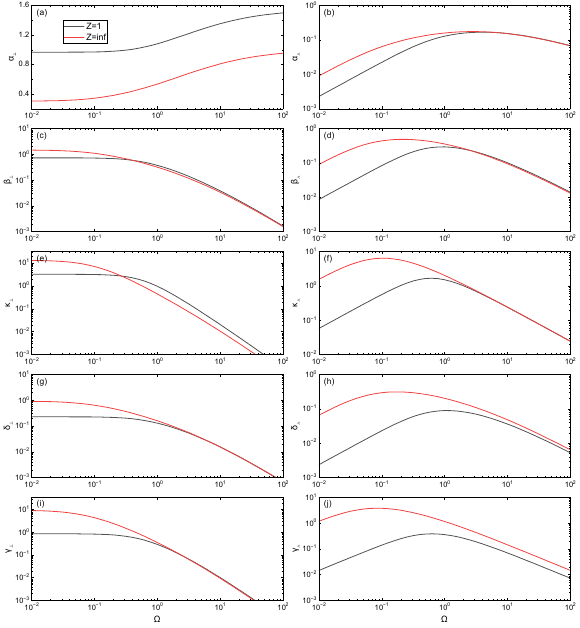}
\caption{Classical transport coefficients from the Fokker-Planck equation. Results for $Z=\infty$ limit is rigorous, while artificial BGK electron-electron collision operator fitted from Fokker-Planck simulations is used for finite Z condition. $\alpha_\perp, \alpha_\wedge$ are the electrical resistivity coefficients. $\beta_\perp, \beta_\wedge$ are the thermoelectric coefficients. $\kappa_\perp, \kappa_\wedge$ are the heat conduction coefficients. $\delta_\perp, \delta_\wedge, \gamma_\perp, \gamma_\wedge$ are the advection coefficients of magnetic field defined in literature \cite{21PRL-Sadler}.}
\label{Fig1}
\end{figure}

In the high-Z limit, $C_{ee}^1(\textbf{f}_1)$ can be neglected and the equation \ref{3b} can be simplified as\cite{06PoP-Nicolai}
\begin{equation}
\label{eq4}
v\nabla f_0 -\frac{e\textbf{E}}{m}\frac{\partial f_0}{\partial v}-\frac{e\textbf{B}}{m}\times\textbf{f}_1=-\nu_{ei}\textbf{f}_1.
\end{equation}
Here $\nu_{ei}=4\pi ne^4Zln\Lambda/mv^3$ is the electron-ion collision frequency. $n$ is the number density of electron, $Z$ is the atomic number, and $ln\Lambda$ is the Coulomb logarithm. In classical Braginskii transport theory, $f_0$ is considered to be the classical Maxwellian EDF, i. e. $f_0^M(\textbf{r},v)=(m/2\pi T)^{3/2}e^{-mv^2/2T}$, here $T$ is the electron temperature. Combining \ref{eq4} with the definition of current\cite{PICF-Atzeni,06PoP-Nicolai}
\begin{equation}
\label{eq5}
J=-\frac{4\pi e}{3}\int_0^\infty \textbf{f}_1v^3dv,
\end{equation}
we can obtain that\cite{84JPDA-Epperlein,06PoP-Nicolai}
\begin{subequations}
\begin{equation}
\label{eq6a}
e\textbf{E}^M=-\frac{\nabla P}{n}+\frac{\textbf{J}\times\textbf{B}}{n}-\beta_{\perp}\nabla T_e-\beta_{\wedge}\times \nabla T+\frac{4m\nu_T}{3\sqrt{\pi}ne}(\alpha_\perp\textbf{J}-\alpha_{\wedge}\textbf{b}\times\textbf{J})
\end{equation}
\begin{equation}
\textbf{f}_1^M = -\frac{1+\textbf{$\chi$}\times}{\nu_{ei}(1+\chi^2)}(v\nabla f_0^M -\frac{e\textbf{E}^M}{m}\frac{\partial f_0}{\partial v})
\end{equation}
\end{subequations}
For clarity we just consider the plane perpendicular to magnetic field here. Equation \ref{eq6a} is the general form of Ohm's law. The first term is the pressure term, the second term is the Hall term, the third and fourth term are the thermoelectric terms, and the last two terms are the electrical resistivity terms. $\nu_T=4\pi ne^4Zln\Lambda/mv_T^3$ is the average electron-ion collision frequency, and $v_T=\sqrt{2T/m}$ is the thermal velocity. $\mathbf{\chi}=\mathbf{\omega}/\nu_{ei}$ is the magnetized parameter and $\mathbf{\omega}=e\textbf{B}/m$ is the electron gyrofrequency. \textbf{b} is the unit vector along the direction of magnetic field. $\beta_{\perp}, \beta_{\wedge}$ are the thermoelectic coefficients, and $\alpha_{\perp}, \alpha_{\wedge}$ are the electrical resistivity coefficients. They can be expressed as the velocity moments of the EDF as\cite{06PoP-Nicolai,22MRE-Zhao}
\begin{subequations}
\begin{equation}
    \beta_\perp = \frac{\phi_3\phi_4+\Omega^2\phi_{4.5}\phi_{5.5}}{\phi_3^2+\Omega^2\phi_{4.5}^2}-\frac{5}{2}
\end{equation}
\begin{equation}
\beta_\wedge = \frac{\phi_3\phi_{5.5}-\phi_{4}\phi_{4.5}}{\phi_3^2+\Omega^2\phi_{4.5}^2}\Omega
\end{equation}
\begin{equation}
\alpha_\perp = \frac{9\pi}{16}\frac{\phi_3}{\phi_3^2+\Omega^2\phi_{4.5}^2}
\end{equation}
\begin{equation}
\alpha_\wedge = \frac{9\pi}{16}(\frac{4}{3\sqrt{\pi}}-\frac{\phi_{4.5}}{\phi_3^2+\Omega^2\phi_{4.5}^2})\Omega
\end{equation}
\end{subequations}
Here $\phi_n = \int_0^\infty d\beta \beta^n e^{-\beta}/(1+\Omega^2\beta^3)$ is the normalized and $\Omega=\omega/\nu_T$ is the magnetized parameter at the thermal velocity.
Using the definition of the electron heat flux $\mathbf{Q}=(2\pi m/3)\int_0^\infty \mathbf{f}_1 v^5 dv$ and the Maxwell equation $\partial\mathbf{B}/\partial t=-\nabla\times\mathbf{E}$ and $\mu_0\mathbf{J}=\nabla\times\mathbf{J}$, we obtain that\cite{21PRL-Sadler}
\begin{subequations}
    \begin{equation}
    \label{eq8a}
        \mathbf{Q}^M = -\frac{3\sqrt{\pi}nT}{4\nu_T}(\kappa_\perp\nabla T+\kappa_\wedge\mathbf{b}\times\nabla T)-\frac{T}{e}(\beta_\perp\mathbf{J}+\beta_\wedge\times\mathbf{J}),
    \end{equation}
    \begin{equation}
    \label{eq8b}
        \frac{\partial\mathbf{B}^M}{\partial t}=\frac{\nabla T\times\nabla n}{ne}+\nabla\times((\mathbf{u}+\mathbf{u}_B)\times\mathbf{B})-\nabla\times(\frac{4m\nu_T}{3\sqrt{\pi}\mu_0ne^2}\alpha_{\perp,\chi=0}\nabla\times\mathbf{B}).
    \end{equation}
\end{subequations}
Equation \ref{eq8a} is the classical Braginskii electron heat flux, $\kappa_\perp, \kappa_\wedge$ are the thermal conductivity coefficients for heat flux along the temperature gradient and perpendicular to both temperature gradient and magnetic field respectively. The equation \ref{eq8b} is the classical extended MHD equation in plasma. At the right hand side, the first term is the Biermann source of magnetic field, the second term is the advection term of magnetic field, and the third term is the resistive diffusion term of magnetic field. $u$ is the plasma fluid velocity, and $u_B$ is the advection velocity of magnetic field, defined as\cite{21PRL-Sadler}
\begin{equation}
\mathbf{u}_B=-(1+\frac{4}{3\sqrt{\pi}}\delta_\perp)\frac{\mathbf{J}}{ne}+\frac{4}{3\sqrt{\pi}}\delta_\wedge \frac{\mathbf{J}\times\mathbf{b}}{ne}-\frac{1}{\nu_T m}(\gamma_\perp\nabla T-\gamma_\wedge\nabla T\times\mathbf{b})
\end{equation}
Here $\delta_\perp = \alpha_\wedge/\Omega, \delta_\wedge = (\alpha_\perp-\alpha_{\perp,\chi=0})/\Omega, \gamma_\perp = \beta_\wedge/\Omega, \gamma_\wedge = (\beta_{\perp,\chi=0}-\beta_\perp)/\Omega$ are the advection coefficients\cite{21PRL-Sadler}. All the theoretical transport coefficients are shown in Fig.~\ref{Fig1}, and are consistent with the simulation results.  Generally, the equation \ref{eq8b} can also be derived through the 3-order moment of equation \ref{eq4}\cite{85PRL-Luciani}
\begin{equation}
\label{eq10}
    \frac{e\mathbf{E}}{m} = \frac{\nabla\int_0^\infty\frac{v^4}{\nu_{ei}}f_0 dv}{\int_0^\infty \frac{v^3}{\nu_{ei}}\frac{\partial f_0}{\partial v}dv}+\frac{\int_0^\infty \frac{v^3}{\nu_{ei}}\mathbf{f}_1 dv}{\int_0^\infty \frac{v^3}{\nu_{ei}}\frac{\partial f_0}{\partial v}dv}\times\frac{e\mathbf{B}}{m}-\frac{3}{4\pi e}\frac{\mathbf{J}}{\int_0^\infty \frac{v^3}{\nu_{ei}}\frac{\partial f_0}{\partial v}dv},
\end{equation}
and the Maxwellian equation $\partial\mathbf{B}/\partial t=-\nabla\times\mathbf{E}$. It is also clear that at the right hand side of the equation \ref{eq10}, the first term is the Biermann source of magnetic field, the second term is the advection term of magnetic field, and the third term is the resistive diffusion term of magnetic field. The form is equivalent to the equation \ref{eq8b} under classical condition $f_0=f_0^M, \mathbf{f}_1=\mathbf{f}_1^M$, and can be extended to the nonlocal condition for general $f_0, \mathbf{f}_1$\cite{02PRL-Kingham}. And the classical transport coefficients are also extended to the nonlocal condition\cite{03PoP-Bychenkov,04PoP-Doumaz}.

At the end of this part, we briefly introduce our method to consider the correction under low-Z condition. To avoid the complexity of directly calculating $C_{ee}^1(\mathbf{f}_1)$ through FP equation, we find an effective BGK collision operator\cite{08PoP-Wallace} as the form $C_{ee}^1(\mathbf{f}_1)=-\nu_{ee}^1\mathbf{f}_1$ can be applicable. Here $\nu_{ee}^1=(\nu_{ei}/Z)\times0.48(v/v_T)^3/(1+0.035(v/v_T)^3)$. The coefficients in the expression are fitted through the FP simulations. Using the collision operator, the low-Z correction $\kappa_\perp(Z)=\kappa_\perp(Z=\infty)\times(Z+0.24)/(Z+4.2)$ and $E^M(Z)=[2/5+3(Z+0.477)/5(Z+2.15))]E^M(Z=\infty)$ at $\Omega=0$ can be recovered. And most of the transport coefficients at low-Z are consistent with the FP simulations\cite{21PRL-Sadler}, except $\alpha_\perp$, which makes no significant influence on the magnetic field advection as $\delta_\wedge = (\alpha_\perp-\alpha_{\perp,\chi=0})/\Omega$ is still relatively accurate\cite{21PRL-Sadler}. Transport coefficients for low-Z condition are also shown in Fig.~\ref{Fig1}.

\section{Nonlocal model}
\subsection{Nonlocal model and corrections of transport coefficients}
We will establish our nonlocal model in this section, especially we will emphasize the importance of considering the correction of electric field under nonlocal condition self-consistently, which is essential for establishing an appropriate nonlocal model.

For convenience, we consider the situation $\Omega=0$ at first.
When the typical wavelength of temperature perturbation is short enough, the departure of zero-order EDF $f_0$ from the Maxwellian EDF $f_0^M$ can't be neglected. Under this condition, we should self-consistently consider the correction $f_0=f_0^M+\Delta f_0, \mathbf{f}_1=\mathbf{f}_1^M+\Delta \mathbf{f}_1,$ and $\mathbf{E}=\mathbf{E}^M+\Delta\mathbf{E}$ and combine \ref{3a}, \ref{3b} together. Substituting the corrections into \ref{3a}, \ref{3b} and neglecting the high-order terms, we can obtain that
\begin{subequations}
    \begin{equation}
    \label{eq11a}
        \frac{1}{v}C_{ee}^0(\Delta f_0)+\nabla\cdot(\frac{\lambda_{ei}}{3}\nabla\Delta f_0)=\frac{1}{3}\nabla\cdot\mathbf{f}_1^M+\nabla\cdot(\frac{e\Delta\mathbf{E}}{3m\nu_{ei}}\frac{\partial f_0^M}{\partial v})
    \end{equation}
    \begin{equation}
    \label{eq11b}
        \Delta\mathbf{f}_1=-\frac{1}{\nu_{ei}}(v\nabla\Delta f_0-\frac{e\Delta\mathbf{E}}{m}\frac{\partial f_0^M}{\partial v})
    \end{equation}
\end{subequations}
Here $\lambda_{ei}=v/\nu_{ei}$ is the electron-ion collision mean free path (mfp). $C_{ee}^0(\Delta f_0)$ usually takes the form of BGK collision operator $-\nu_{ee}\Delta f_0$ or the AWBS collision operator $\nu_{ee}v(\partial \Delta f_0/\partial v)$. $\nu_{ee}=r\nu_{ei}/Z$ is the effective electron-electron collision frequency and can be adapted through a constant $r$\cite{17PoP-Brodrick}. $\Delta\mathbf{E}$ needs to be determined from the the equation \ref{eq5}. Substituting the corrections into equation \ref{eq5} (current balance of $\Delta\mathbf{f}_1$), we obtain
\begin{equation}
\label{eq12}
\frac{e\Delta\mathbf{E}}{m} = \frac{\nabla\int_0^\infty\frac{v^4}{\nu_{ei}}\Delta f_0dv}{\int_0^\infty \frac{v^3}{\nu_{ei}}\frac{\partial f_0^M}{\partial v}dv}
\end{equation}
Through solving equations \ref{eq11a} and \ref{eq12}, which is convenient in RHD code through the multi-group method\cite{00PoP-Schurtz}, we can obtain the corrections of $f_0$ and $\mathbf{E}$, and then using equation \ref{eq11b} to derive $\Delta\mathbf{f}_1$. Using the corrections of EDF and electric field, we can obtain the corrections of heat flux $\mathbf{q}_e$, Biermann source $\partial\mathbf{B}/\partial t$, and Nernst velocity $\mathbf{v}_N$\cite{85PRL-Luciani}:
\begin{subequations}
    \begin{equation}
    \label{eq13a}
        \mathbf{q}_e = \frac{2\pi m}{3}\int_0^\infty(\mathbf{f}_1^M+\Delta\mathbf{f}_1) v^5dv,
    \end{equation}
    \begin{equation}
    \label{eq13b}
        \frac{e}{m}\frac{\partial\mathbf{B}}{\partial t}=-\nabla\times\frac{\nabla\int_0^\infty\frac{v^4}{\nu_{ei}}(f_0^M+\Delta f_0)dv}{\int_0^\infty \frac{v^3}{\nu_{ei}}\frac{\partial (f_0^M+\Delta f_0)}{\partial v}dv},
    \end{equation}
    \begin{equation}
        \label{eq13c}
        \mathbf{v}_N = -\frac{\int_0^\infty \frac{v^3}{\nu_{ei}}(\mathbf{f}_1^M+\Delta\mathbf{f}_1)dv}{\int_0^\infty \frac{v^3}{\nu_{ei}}\frac{\partial f_0}{\partial v}dv}.
    \end{equation}
\end{subequations}
Following this process we can obtain nonlocal electron heat flux and electric field self-consistenly. For comparison, in SNB model\cite{00PoP-Schurtz}, which is typically one of the most popular nonlocal models, and its improvement, $\Delta\mathbf{E}$ is not self-consistently involved in the equations of nonlocal correction. Only an equation is considered to solve $\Delta f_0$\cite{00PoP-Schurtz}
\begin{equation}
\label{eq14}
    \frac{1}{v}C_{ee}(\Delta f_0)+\nabla\cdot(\frac{\lambda_{ei}}{3}\nabla\Delta f_0)=\frac{1}{3}\nabla\cdot\mathbf{f}_1^M
\end{equation}
In some literature, $\mathbf{f}_1^M$ can also be artificially modified to a different form as $\mathbf{g}_1^M$ for improvement\cite{00PoP-Schurtz,06PoP-Nicolai,17PoP-Brodrick,25Arxiv-Chen}. The main difference of SNB model from our model is that $\Delta f_0$ and $\Delta\mathbf{E}$ are not self-consistently determined, but using $\Delta f_0$ to determine $\Delta\mathbf{E}$ through equation \ref{eq12} (or equation \ref{eq13b} equivalently) directly. And $\Delta\mathbf{f}_1$ is also determined with a different form\cite{00PoP-Schurtz}
\begin{equation}
    \Delta\mathbf{f}_1=-\frac{1}{\nu_{ei}}(v\nabla\Delta f_0).
\end{equation}
Using the corrections of EDF and equations \ref{eq13a}, \ref{eq13b}, \ref{eq13c}, one can also determine the corrections of electron heat flux and electric field. But we will show that neither $\mathbf{f}_1^M$ nor $\mathbf{g}_1^M$ gives the reasonable results.

\begin{figure*}[htbp]
\centering
\includegraphics[width=1.0\textwidth]{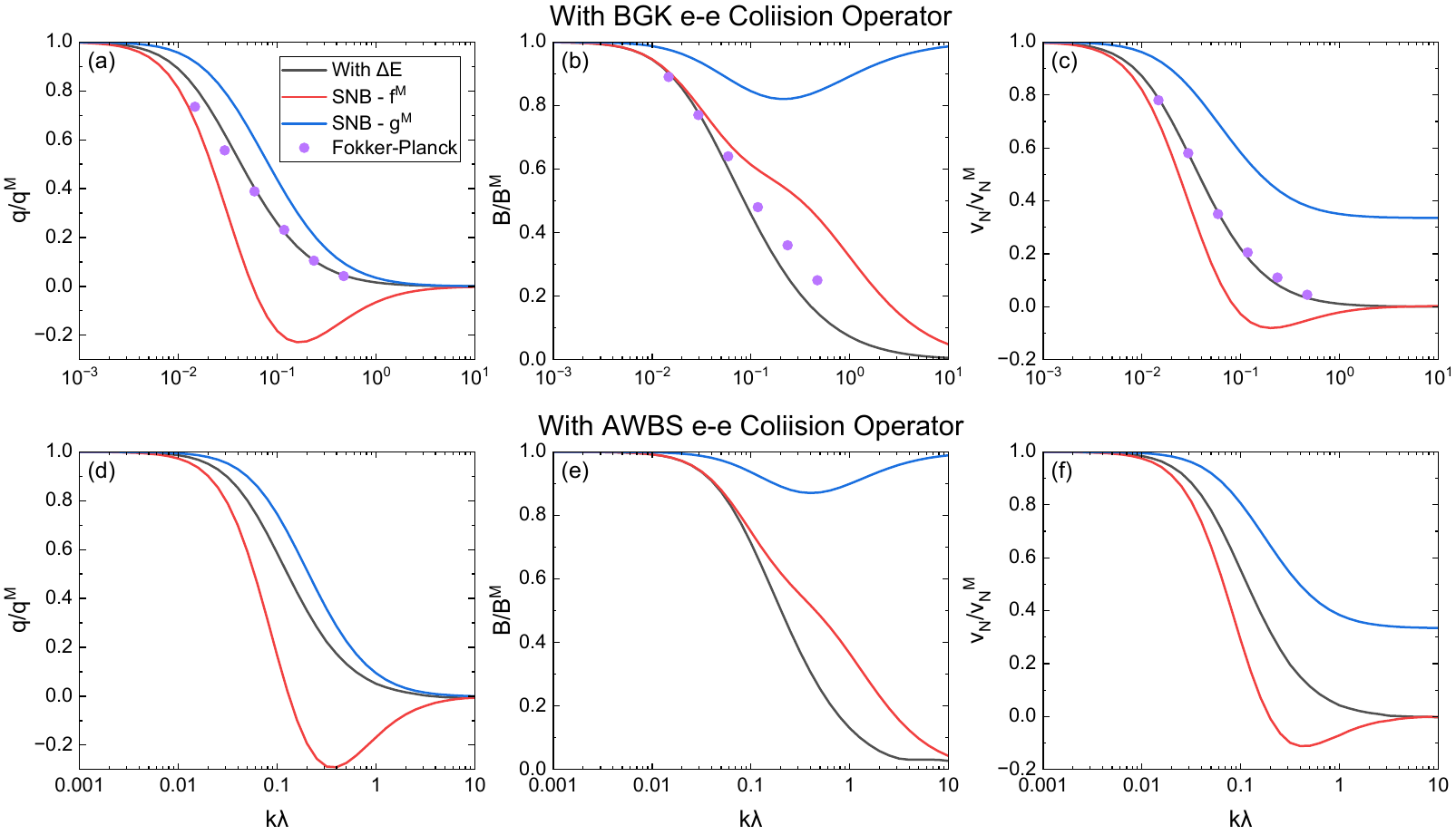}
\caption{Normalized nonlocal electron heat flux, Biermann magnetic, and Nernst velocity with small temperature perturbation in unmagnetized plasma from our model, SNB model with $\mathbf{f}_1^M$ and $\mathbf{g}_1^M$ source terms respectively. The circles are simulation results from Fokker-Planck code IMPACTA\cite{04JCP-Kingham} corresponding to $r=2$ in nonlocal model. Results from BGK and AWBS electron-electron collision operators are both shown.}
\label{Fig2}
\end{figure*}

Using the classical test\cite{24PPCF-Yuan} of small perturbation of density and temperature $T=T_0(1+\epsilon_1 \cos k_1x), n=n_0(1+\epsilon_2 \cos k_2y)$, we can derive the corrections of the above quantities through different nonlocal models to take the examination. Here $T_0, n_0$ are the average electron temperature and number density, and $\epsilon_1, \epsilon_2<<1$ are the relative perturbation amplitude. $k_1, k_2$ are the wave number of the perturbation along the x and y directions. We take $k_2<<k_1$ for convenience. The results of the corrections of electron heat flux, Biermann source and Nernst velocity normalized by their classical local values at different $k\lambda$ are shown in the Fig.~\ref{Fig2}, here $k$ refers to $k_1$ and $\lambda=\sqrt{\lambda_{ei}\lambda_{ee}}$ is the effective average mfp at the thermal velocity. Calculation with different collision operators and nonlocal models are all carried.

As shown in Fig.~\ref{Fig2}, the results from BGK and AWBS electron-electron collision operator are similar. However, results from different model are significantly different. Our model can give the nonlocal suppression for electron heat flux, Biermann source and Nernst velocity together. When $k\lambda$ increases from $0$ to $\infty$, the normalized nonlocal quantities are all suppressed from $1$ to $0$\cite{24PPCF-Yuan,23PoP-Davies,24PoP-Walsh}. We also show the Fokker-Planck simulation results through the code IMPACTA\cite{04JCP-Kingham} along with the BGK results. Here $r=2$ is used. The Fokker-Planck simulation results are consistent with our nonlocal model for all quantities. However, neither of the SNB models can recover the correct trend.  Taking $\mathbf{f}_1^M$ as the source term of equation \ref{eq14}, the quantities can be suppressed at large $k\lambda$. But the electron heat flux $q$ and Nernst velocity $v_N$ turn negative with $k\lambda>0.1$, which violates the second law of thermodynamics. If taking $\mathbf{g}_1^M$ as the source term in equation \ref{eq14}, suppression of $q$ is appropriate, but nonlocal suppression of Biermann source and Nernst velocity are not significant. The reason for the unreasonable results of SNB model is that the correction of electric field in equation \ref{eq12} (equivalent to \ref{eq5}) is not self-consistently considered, which means the constraint of current is violated. So the return current\cite{PICF-Atzeni} is not self-consistently included in the SNB model, resulting in the unreasonable corrections of $\mathbf{f}_1$, and then the electron heat flux and Nernst velocity. As for $\mathbf{g}_1^M$, it is an artificial adaptation to match the heat flux, but it can't recover the suppression of $B$ and $v_N$ self-consistently. This can be clearly shown through the analytic solution of the nonlocal corrections.

Specifically, we can derive the analytic solution of the nonlocal corrections with BGK operator. For our model, the nonlocal corrections for heat flux and Nernst velocity are
\begin{subequations}
\label{eq16}
\begin{equation}
\label{eq16a}
    \frac{q}{q^M} = 1-\frac{1}{24}\int_0^\infty \frac{k^2\lambda^2\beta^4/3}{1+k^2\lambda^2\beta^4/3}\beta^4(\beta-4-\frac{3}{k^2\lambda^2\beta^4}\epsilon)e^{-\beta}d\beta
\end{equation}
\begin{equation}
    \frac{v_N}{v_N^M} = 1-\frac{64}{2835\sqrt{\pi}}\int_0^\infty \frac{k^2\lambda^2\beta^4/3}{1+k^2\lambda^2\beta^4/3}\beta^{\frac{9}{2}}(\beta-4-\frac{3}{k^2\lambda^2\beta^4}\epsilon)e^{-\beta}d\beta
\end{equation}
\end{subequations}
Here $\epsilon = \Delta E/(k_1 \epsilon_1 T_0)=\int_0^\infty [k^2\lambda^2\beta^4/3/(1+k^2\lambda^2\beta^4/3)]\beta^3(\beta-4)e^{-\beta}d\beta/\int_0^\infty \beta^3 e^{-\beta}/(1+k^2\lambda^2\beta^4/3)d\beta\sim\Delta E/E^M$ is the normalized electrical field correction. $\epsilon$ increases with $k\lambda$ and $\epsilon\sim 1$ even at $k\lambda\sim 0.1$, so the correction of electric field is significant.
For comparison, in SNB model, the nonlocal corrections are
\begin{subequations}
\begin{equation}
\label{eq17a}
    \frac{q}{q^M} = 1-\frac{1}{24}\int_0^\infty \frac{k^2\lambda^2\beta^4/3}{1+k^2\lambda^2\beta^4/3}\beta^4(\beta-4)e^{-\beta}d\beta
\end{equation}
\begin{equation}
    \frac{v_N}{v_N^M} = 1-\frac{64}{2835\sqrt{\pi}}\int_0^\infty \frac{k^2\lambda^2\beta^4/3}{1+k^2\lambda^2\beta^4/3}\beta^{\frac{9}{2}}(\beta-4)e^{-\beta}d\beta
\end{equation}
\end{subequations}
The results corresponds to the $\mathbf{f}_1^M$ source term, the term $(\beta-4)$ in these expressions vanish for the $\mathbf{g}_1^M$ source term. Clearly the corrections of electric field take no effects in the expressions. Analytic nonlocal corrections of Biermann magnetic field can also be obtained for different models. The corrections are essentially 2-dimension effects and rather complex, including different order velocity moments of $\Delta f_0$. They will be shown along with the detailed process to get these analytic solutions in a more detailed edition the future.

Comparing the equation \ref{eq16a} and \ref{eq17a}, the normalized electric field corrections $\epsilon$ contributes to the term involving return current in the heat flux correction $\Delta q$, it reduces $\Delta q$ to smaller than 1, avoiding the negative nonlocal $q$. However, electric field corrections are not involved in the SNB model, and related return current doesn't take effects, thus the correction $\Delta q$ gets larger than $1$ unphysically, resulting in negative heat flux. As for the $\mathbf{g}_1^M$ source term, it's just an artificial adaptation considering $\int_0^\infty \beta^4d\beta=\int_0^\infty \beta^4(\beta-4)d\beta$\cite{00PoP-Schurtz}, so it can't recover the suppression of other quantities which involving high-order velocity moments. This analysis clearly shows the effects of considering corrections of electric field self-consistently.

Then we extend the nonlocal model to magnetized plasma. The equations to solve the corrections of EDF and electric field are
\begin{subequations}
    \begin{equation}
    \label{eq18a}
        \frac{1}{v}C_{ee}^0(\Delta f_0)+\nabla\cdot(\frac{\lambda_{ei}(1+\mathbf{\chi}\times)}{3(1+\chi^2)}\nabla\Delta f_0)=\frac{1}{3}\nabla\cdot\mathbf{f}_1^M+\nabla\cdot(\frac{1+\mathbf{\chi}\times}{1+\chi^2})(\frac{e\Delta\mathbf{E}}{3m\nu_{ei}}\frac{\partial f_0^M}{\partial v})
    \end{equation}
    \begin{equation}
    \begin{aligned}
        \frac{e\Delta\mathbf{E}}{m} = \frac{\int_0^\infty \frac{v^3}{\nu_{ei}(1+\chi^2)}\frac{\partial f_0^M}{\partial v}dv-\int_0^\infty \frac{v^3\mathbf{\chi}}{\nu_{ei}(1+\chi^2)}\frac{\partial f_0^M}{\partial v}dv\times}{(\int_0^\infty \frac{v^3}{\nu_{ei}(1+\chi^2)}\frac{\partial f_0^M}{\partial v}dv)^2+(\int_0^\infty \frac{v^3\chi}{\nu_{ei}(1+\chi^2)}\frac{\partial f_0^M}{\partial v}dv)^2}\\(\int_0^\infty\frac{v^4}{\nu_{ei}(1+\chi^2)}\nabla\Delta f_0dv+\int_0^\infty\frac{v^4\mathbf{\chi}\times}{\nu_{ei}(1+\chi^2)}\nabla\Delta f_0dv)
    \end{aligned}
    \end{equation}
\end{subequations}
And $\Delta\mathbf{f}_1$ can be determined through
\begin{equation}
    \Delta\textbf{f}_1^M = -\frac{1+\textbf{$\chi$}\times}{\nu_{ei}(1+\chi^2)}(v\nabla\Delta f_0 -\frac{e\Delta\textbf{E}}{m}\frac{\partial f_0}{\partial v})
\end{equation}
Through these corrections, nonlocal corrections of electron heat flux and magnetic field can also be obtained for magnetized plasma.

\begin{figure*}[htbp]
\centering
\includegraphics[width=1.0\textwidth]{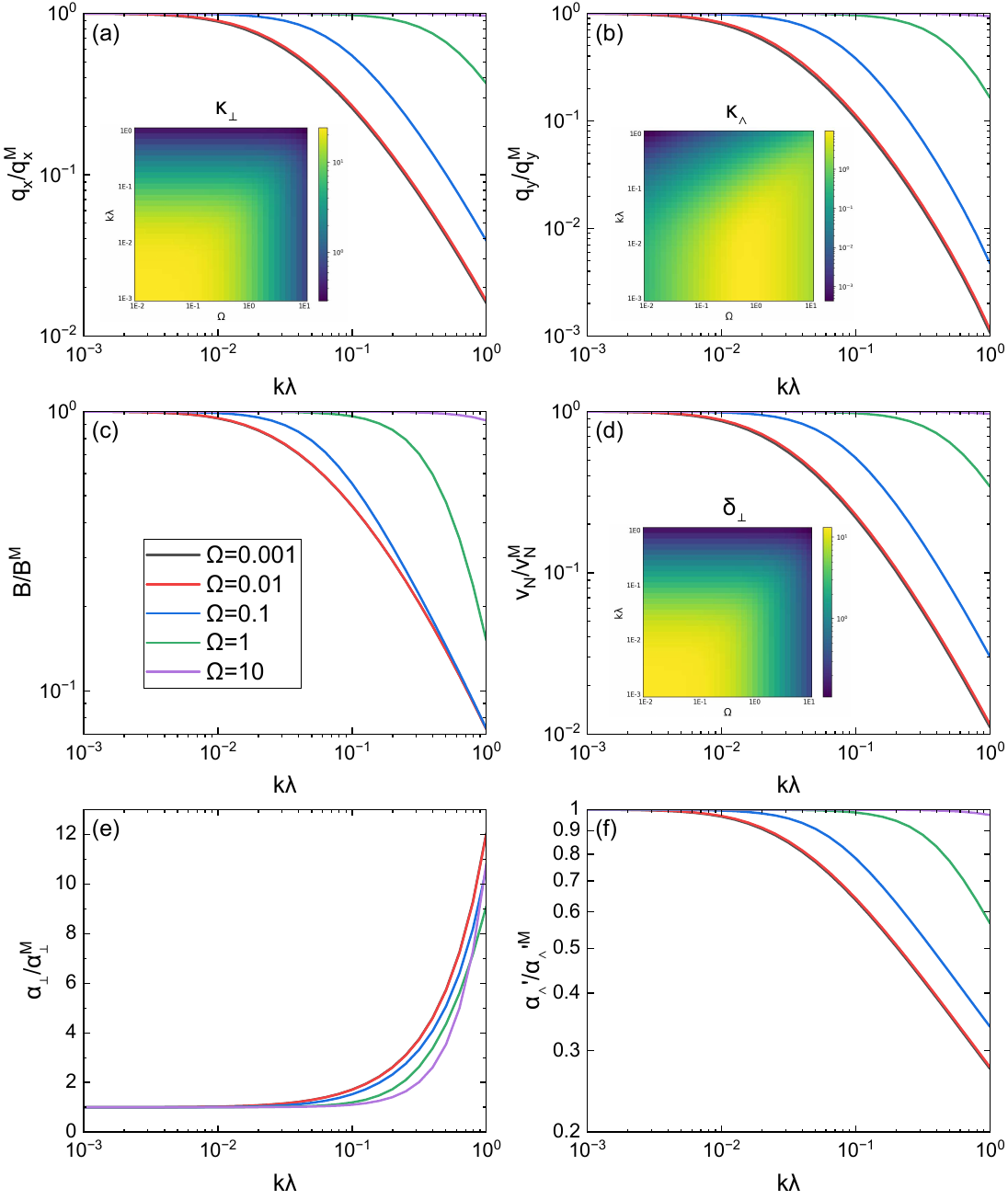}
\caption{Normalized nonlocal results of heat flux $q_x$, $q_y$, Biermann magnetic field, Nernst velocity, and electric resistivity $\alpha_\perp, \alpha_\wedge'$ with small temperature perturbation and different magnetized parameters. Here $\alpha_\wedge'=\alpha_\wedge-3\sqrt{\pi}\Omega/4$ refers to the total coefficient of transverse electric field $E_y$ (including Hall effects) corresponding to current $j_x$.}
\label{Fig3}
\end{figure*}

Similarly, we can get the nonlocal correction in the classical test of small perturbation of density and temperature $T=T_0(1+\epsilon_1 \cos k_1x), n=n_0(1+\epsilon_2 \cos k_2y), B=B_0$, as a typical example to show how the magnetic field affects the nonlocal effects for different coefficients. The results of the corrections of electron heat flux along x and y directions (Righi-Leduc heat flux), Biermann source and Nernst velocity at different $k\lambda$ and magnetized parameters normalized by their classical local values are shown in the Fig.~\ref{Fig3}. Cross-field Nernst velocity is not shown here as it takes no effect in this geometry. BGK collision operator is used for convenience.

As shown in Fig.~\ref{Fig3}, the electron heat flux along x and y directions, Biermann magnetic field, and Nernst velocity are all suppressed by the nonlocal effects at all magnetized parameters. When $k\lambda$ increases from $0$ to $\infty$, the normalized values are all suppressed from $1$ to $0$. Furthermore, the effects of nonlocal suppression is reduced at larger magnetized field. Because the radius of gyration of electrons in magnetic field decreases with larger magnetic field, which is equivalent to the decrease of mfp, thus reducing the delocalization of the electrons. It can be seen from the equation \ref{eq18a} that the effective mfp is reduced to approximate $\lambda^*\sim\lambda/(1+\Omega^2)$, so the nonlocal effects at larger $\Omega$ is less significant. We also find that for different $k\lambda$, dependence of these transport coefficients on magnetized parameter are different. For $q_{x}$ ($\kappa_\perp$) and $v_N$, the peak values are still at $\Omega=0$ and they decrease with increasing $\Omega$ at all $k\lambda$. But the decrease is less significant at larger $k\lambda$ and they get almost constant for large $\Omega$. As for $q_{y}$ ($\kappa_\wedge$), $\Omega$ where it gets the peak value shifts. We also examine the SNB model with different source terms, including original $\mathbf{f}_1^M, \mathbf{g}_1^M$ and an improved form of $\mathbf{g}_1^M$\cite{00PoP-Schurtz,06PoP-Nicolai,17PoP-Brodrick,25Arxiv-Chen} respectively in magnetized plasma, both $q_x$ and $q_y$ turn negative in certain range of $\Omega$ and $k\lambda$. This indicates that in more complex situation, it's more difficult to get reasonable quantities without self-consistently considering electric field correction rigorously.

We also give simple discussion of nonlocal correction of the electrical resistivity coefficients. Electrical resistivity tensors under nonlocal condition is rather complex, we just consider a simplified condition that $j_x=j_0 \cos kx$ and $j_y=0$, i.e. the direction of perturbation wave number is along the current. Normalized nonlocal corrections of $\alpha_\perp$ and $\alpha_\wedge$ at different magnetized parameters are shown in Fig.~\ref{Fig3}. Unlike other transport coefficients, $\alpha_\perp$ increases with the increasing nonlocal parameter $k\lambda$\cite{04PoP-Doumaz}, which means larger electric field is needed to drive the current. As for $\alpha_\wedge$, the transverse electrical resistivity term itself isn't  suppressed from $1$ to $0$ with increasing $k\lambda$ from $0$ to $\infty$. The combination of transverse electrical resistivity term and the Hall term $\alpha_\wedge-3\sqrt{\pi}\Omega/4$\cite{06PoP-Nicolai,24PoP-Lopez}, i.e. the total coefficient of transverse electric field $E_y$, is suppressed from $1$ to $0$ with increasing $k\lambda$. And larger magnetized parameters reduce the nonlocal effects.

\subsection{Magnetic field generation without density gradients}
We study the nonlocal magnetic field generation without density gradients\cite{02PRL-Kingham,04JCP-Kingham} using the nonlocal model. Classical Biermann magnetic field generation is $\partial\mathbf{B}/\partial t=(\nabla T\times\nabla n)/ne$. Self-generated magnetic field is $0$ without density gradients. As shown in equation \ref{eq10} and \ref{eq13b}, generally the magnetic field is generated through
\begin{equation}
\label{eq20}
    \frac{e}{m}\frac{\partial\mathbf{B}}{\partial t}=\frac{\nabla\int_0^\infty \frac{v^3}{\nu_{ei}}\frac{\partial f_0}{\partial v}dv\times\nabla\int_0^\infty\frac{v^4}{\nu_{ei}}f_0dv}{(\int_0^\infty \frac{v^3}{\nu_{ei}}\frac{\partial f_0}{\partial v}dv)^2},
\end{equation}
In classical local limit $f_0=f_0^M$ and the expression is reduced to the classical Biermann source. When the electron number density is constant, the two velocity moments of $f_0^M$ in \ref{eq20} are both the functions of $T$ so the cross product is $0$, resulting classical Biermann source to be $0$. However, generally $f_0\neq f_0^M$ so self-generated magnetic field can also exist even without density gradients\cite{02PRL-Kingham,04JCP-Kingham}, especially under nonlocal condition where departure of $f_0$ from $f_0^M$ is significant. This mechanism has been proven through Fokker-Planck simulations\cite{02PRL-Kingham,04JCP-Kingham}. We study the nonlocal magnetic field through our model and find it originates from the different typical perturbation wave number in orthogonal directions and widely exists.

\begin{figure*}[htbp]
\centering
\includegraphics[width=1.0\textwidth]{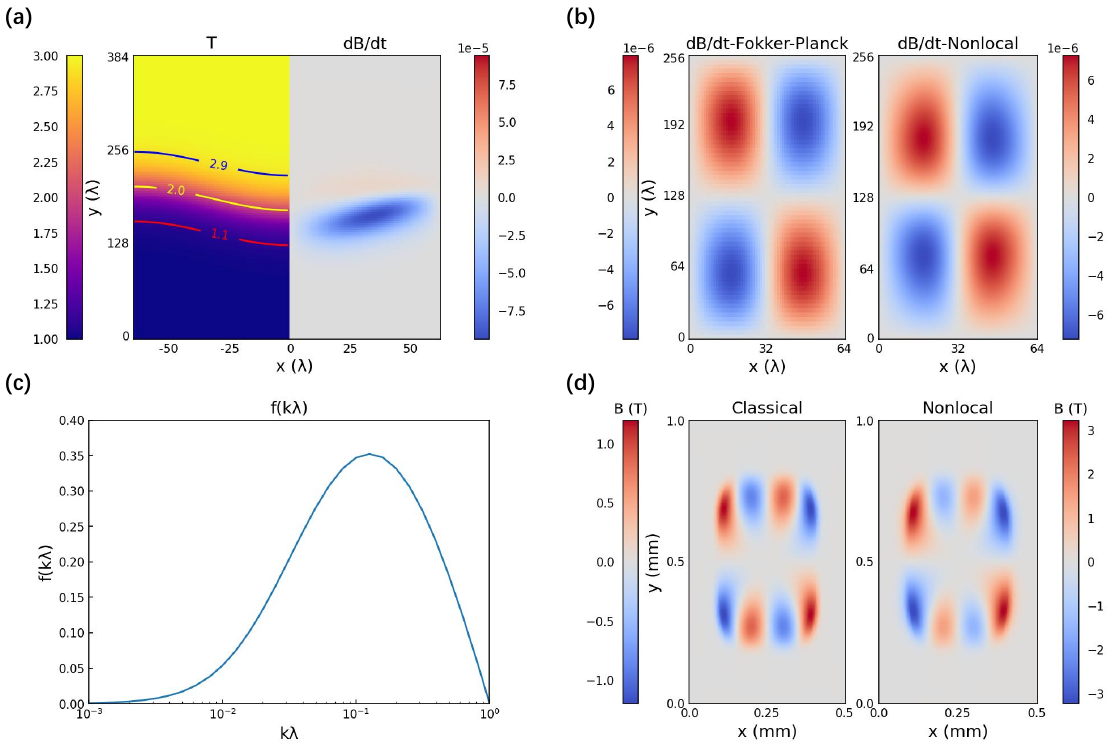}
\caption{Results for magnetic field generation without density gradients. (a)Normalized temperature distribution and magnetic field generation with normalized temperature distribution $T=2+tanh[(x-192-16\cos(2\pi y/128))]$. (b)Normalized magnetic field generation with small temperature perturbation from Fokker-Planck simulations and our model. (c)The coefficient $f(k\lambda)$ for nonlocal magnetic field generation with small temperature perturbation in a wide range. (d)Magnetic field at 0.3 ns through FLASH simulations\cite{00-FLASH}. The results are from only Biermann source and with nonlocal corrections.}
\label{Fig4}
\end{figure*}

Fig.~\ref{Fig4} (a) shows the normalized $\partial B_z/\partial t$ calculated by our model with the temperature distribution $T=T_0(2+tanh[(x-Lx/2-Ly/8\times \cos(2\pi y/Ly))/(Lx/12)])$ and constant electron number density $n_0$. Here $0<x<L_x, 0<y<L_y$ are the coordinates normalized by mfp $\lambda_0$. $L_x=384\lambda_0$ and $L_y=128\lambda_0$ are the size of the computational domain. $\lambda_0$ is the mfp with temperature $T_0$. Magnetic field is mainly generated in the region with large temperature gradient, consistent with the results of Fokker-Planck simulations. The temperature distribution is typical distribution at the ablation front in ICF with laser imprint effects, so the nonlocal magnetic field generation without density gradients exists along with classical Biermann source at the ablation front.

To analysis the reason for the mechanism physically, we use the temperature distribution $T=T_0(1+\epsilon_1\cos k_1x)(1+\epsilon_2\cos k_2y)$ and calculate the corresponding magnetic field generation. Fig.~\ref{Fig4} (b) shows the results for $k_1\lambda=2\pi/64, k_2\lambda=2\pi/256,$ and $ \epsilon_1=\epsilon_2=0.1$ from our model and Fokker-Planck simulations. The distribution of magnetic field calculated by our model is consistent with the Fokker-Planck simulation and the values are close. The spatial distribution is $B\propto\sin k_1x\sin k_2y$ in this case. With different $k_1$ and $k_2$, we find the magnitude of the magnetic field generation is
\begin{equation}
    \frac{\partial B}{\partial t}=[f(k_1\lambda)-f(k_2\lambda)]k_1\epsilon_1 k_2\epsilon_2 T_0 
\end{equation}
Here $f(k\lambda)$ is the function of nonlocal parameter $k\lambda$ as shown in Fig.~\ref{Fig4} (c) and $f(k\lambda)\propto(k\lambda)^2$ when $k\lambda<<1$. This can also be written as $\partial B/\partial t\propto[f(k_1\lambda)-f(k_2\lambda)](\nabla_x T\times\nabla_y T)/T$ and $\partial B/\partial t\propto[(k_1\lambda)^2-(k_2\lambda)^2](\nabla_x T\times\nabla_y T)/T$ when $k\lambda<<1$. Through the expression, we find that the magnetic field generation without density gradient is from the cross product of the temperature gradients along the orthogonal directions. And it's more significant when the perturbation wavelength along the orthogonal directions are different. If $k_1=k_2$ the generation vanish. The generation is more significant when nonlocal effects are more significant. So the generation originates from the different nonlocal effects along orthogonal directions. Physically, this is because the nonlocal corrections of electric field are different along orthogonal directions with different $k\lambda$, which contributes to the net curl of the electric field, resulting in magnetic field generation according to $\partial\mathbf{B}/\partial t=-\nabla\times\mathbf{E}$ even without density gradients. This can be more clearly by transform the above expression into the form $\partial B/\partial t\propto k_2\epsilon_2\Delta E_x-k_1\epsilon_1\Delta E_y$  according to equations \ref{eq16} that $\Delta E\propto(k\lambda)^2\nabla T$ while $k\lambda<<1$. It is clear that the magnetic field originates from the curl of the nonlocal correction of electric field. As the perturbation wavelength are different along the different directions, the magnetic field generation without density gradient exists. The mechanism takes no effects under classical condition as $\Delta E=0$. Considering $\nabla\delta T\times\nabla(\nabla^2\delta T)=(k_2^2-k_1^2)k_1\epsilon_1 k_2\epsilon_2 T^2 \sin k_1x\sin k_2y$ for the temperature distribution, the expressions can also be written as the form $\partial B/\partial t\propto \nabla\delta T\times\nabla(\nabla^2\delta T)$ in this case while $k\lambda<<1$, consistent with the literature\cite{02PRL-Kingham,04JCP-Kingham}.

We show the effects of this mechanism in comprehensive RHD simulations. The nonlocal model is implemented into the code FLASH to calculate the nonlocal effects. We consider the initial electron temperature to be $T=T_0 exp(-(x/r_x)^4-(y/r_y)^4)$ and the initial electron number density to be $n=n_0$. Here $T_0=1 $ keV and $n_0=3\times10^{21}$  $cm^{-3}$. The average radius $r_x, r_y$ are $250\mu m$ and $125\mu m$ respectively, corresponding to the typical nonlocal parameter $\lambda/L_T~0.01$. This is the typical distribution after laser irradiates the gas target. During the first 0.3 ns of the simulation, electrons are pushed around by the higher central pressure due to the initial temperature distribution, resulting in nearly parallel temperature gradient and density gradient so the Biermann magnetic field is relatively small. Magnetic field generation without density gradients can be highlighted in this geometry with $r_x\neq r_y$. Fig.~\ref{Fig4} (d) shows the spatial distribution of magnetic field at 0.3 ns from FLASH simulation. The classical condition just considers the Biermann source while the nonlocal condition also considers the nonlocal corrections. Spatial distribution of magnetic field are similar between the two conditions. Under classical condition, the peak value is about $1$ T, while it increases to about $3$ T under the nonlocal condition. Nonlocal corrections additionally contribute about $2$ T to the magnetic field in this case mainly through the mechanism without density, which indicates that nonlocal corrections of magnetic field can be significant under this typical condition. This reveals that besides the suppression of classical Biermann source, the nonlocal effects can also amplify the magnetic field though mechanism vanishing in classical condition. We also find the nonlocal correction takes no effects in cylindrical symmetry geometry in simulation as we interpreted above.

\subsection{Nonlocal effects in laser ablation}
\begin{figure*}[htbp]
\centering
\includegraphics[width=1.0\textwidth]{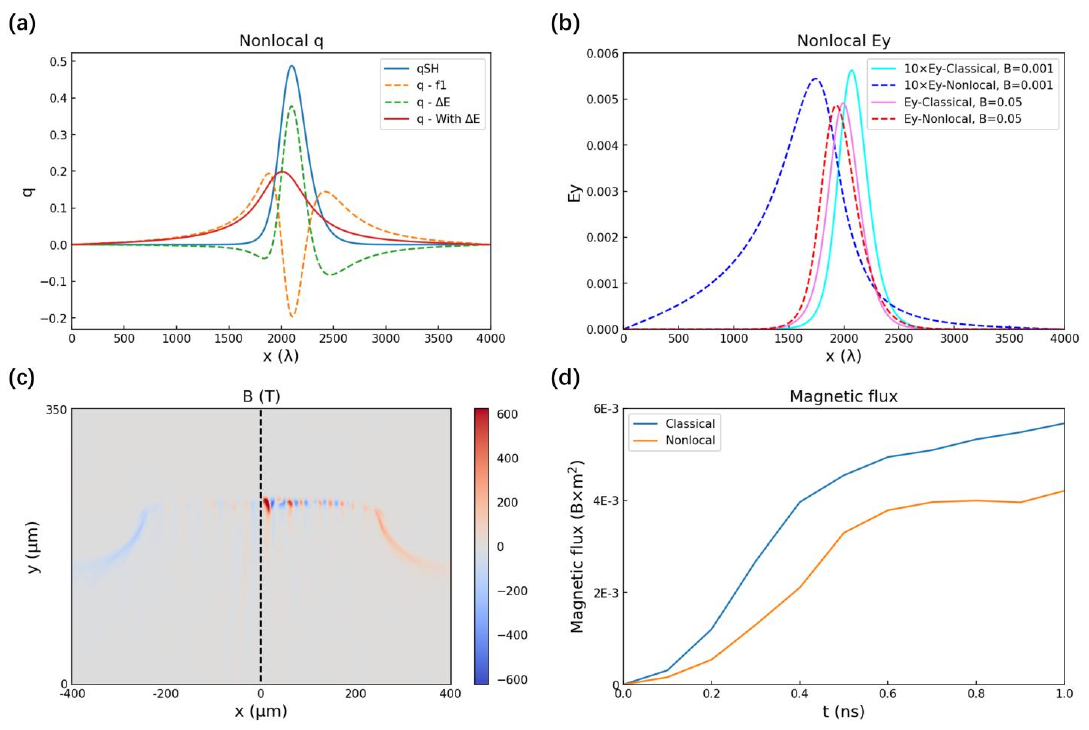}
\caption{(a)Normalized electron heat flux with steep temperature gradient in unmagnetized plasma. $qSH$ refers to the classical Spitzer-Harm heat flux. $q-f1$ refers to the nonlocal heat flux without self-consistently considering electric field correction (i.e. the nonlocal heat flux in SNB model with $\mathbf{f}_1^M$ source term). $q-\Delta E$ refers to the heat flux corresponding to the return current from electric field correction. $q-With \Delta E$ refers to the total nonlocal heat flux in our model. (b)Normalized classical and nonlocal transverse electric field with steep temperature gradient with different magnetized parameters. The results with $\Omega=0.001$ is artificially enhanced by 10 times to be more clear. (c)Magnetic field at 1.0 ns in laser ablation from FLASH simulations. The left part is the results from classical Biermann source and the right part is with the nonlocal corrections. (d)Magnetic flux in half the space.}
\label{Fig5}
\end{figure*}

We use the nonlocal model to study heat flux and Nernst velocity suppression with large temperature gradients. As an example, the normalized electron heat flux with the temperature distribution $T=T_0(3+2tanh[(x-2000\lambda_0)/200\lambda_0])$, which is typical temperature distribution in laser ablation, is shown in Fig.~\ref{Fig5} (a). Electron number density is constant and magnetic field is $0$. Here $\lambda_0$ is the mfp at temperature $T_0$, with typical nonlocal parameter $\lambda/L_T\sim0.1$ in this case. Suppression of the peak heat flux and preheating in the cold region are significant with the parameters. The heat flux originates from $\Delta f_0$ in equation \ref{eq11b} (i.e. the heat flux with source term $\mathbf{f}_1^M$ in SNB model) and $\Delta E$ are specifically shown respectively. It is clear that if the correction of electric field is neglected, the nonlocal heat flux turns negative at the location of peak flux. The correction of electric field drives positive heat flux corresponding to the return current, compensating for the negative heat flux, consistent with our analysis under small perturbation condition. With the combination of contribution from $\Delta\mathbf{E}$ and $\Delta f_0$, the total heat flux gets appropriate value. This shows the importance of self-consistently considering the correction of electric field corresponding to the current balance of $\Delta \mathbf{f}_1$ again. The normalized electric field along the y direction $E_y$ (i.e. $\mathbf{v}_N\times\mathbf{B}$) under the same temperature distribution is shown in Fig.~\ref{Fig5} (b) with the normalized magnetic field $B = 0.001$ and $0.05$ respectively. The corresponding magnetized parameter is approximate $\Omega<0.01$ and $\Omega~0.2$ at the location with largest $E_y$. The value of $E_y$ is multiplied by 10 times at $B=0.001$ as its original value is relatively small. The peak value is not suppressed significantly in this case for both of the parameters. We also find the suppression is more significant with steeper temperature gradient. The location of peak value shifts toward the cold region, consistent with previous literature\cite{18PPCF-Brodrick}. With larger magnetic field, the change of $E_y$ from nonlocal effects decreases as stronger magnetization confines the eletron movement and reduces the effective mfp.

Finally we show the simulation results of magnetic field in laser ablation of CH target with nonlocal model implemented in FLASH. The spatial profile of the laser is super-Gaussian with exponent 4, radius $200 \mu m$, and constant peak intensity $I=2\times 10^{15} W/cm^2$. The CH target is with initial density $1.05$ g/cc and intinial temperature $1$ eV. Magnetic field at 1 ns with only classical Biermann source and with nonlocal correction are shown in Fig.~\ref{Fig5} (c). For both conditions, magnetic field mainly concentrates at the edge of the laser spot on the surface of the target, where density and temperature gradients are largest. Nonlocal effects significantly change the spatial distribution and peak value of the magnetic field. The area where magnetic field concentrate at significantly shrinks especially in corona, because nonlocal effects suppress the Biermann generation. The suppression is more significant in low-density corona because mfp is larger. Magnetic flux in half the area is suppressed to about $2/3$ times with nonlocal corrections as shown in Fig.~\ref{Fig5} (d). However, as for peak value, the condition is more complex. Under classical condition, the peak magnetic field is at the edge of laser spot on target surface. With nonlocal correction, value of magnetic field in this area doesn't change significantly. But the high-mode magnetic field perturbation in the center of the laser spot significantly increases, whose peak amplitude can even be larger than the magnetic field at the edge. We attribute this significant magnetic field perturbation to the high-mode transverse temperature perturbation from laser imprint and hydrodynamic instabilities. The transeverse temperature perturbation, together with the steep longitudinal temperature gradient at ablation front, generates nonlocal magnetic field according to the mechanism interpreted above. In this case, the nonlocal effects on magnetic field are complex. Besides the global suppression of Biermann source, local magnetic field perturbation can be even larger with non-classical mechanism, which may have complex affections on the evolution of hydrodynamic instabilities\cite{22PoP-Zhang,24JFM-Zhang}.

\section{Conclusion}

We propose a new nonlocal model which can simultaneously recover the nonlocal corrections of heat flux and magnetic field in RHD simulations. We clarify the necessity of considering the electric field correction self-consistently to establish an appropriate nonlocal model through the comparison of our model and SNB models. Using the nonlocal model, we study the nonlocal corrections of classical transport coefficients in magnetized plasma and systematically study the non-classical magnetic field generation mechanism without density gradients. We also find the complex influence of the nonlocal effects on magnetic field distribution in laser ablation, which potentially influences the hydrodynamic instabilities. We establish an self-consistent nonlocal model for the magnetic field evolution for the first time and reveals the complex nonlocal effects in ICF.

Implementation of the nonlocal model in RHD code still needs development in detail, we will develop the implementation and use comprehensive RHD simulations to study the nonlocal effects for coupling of magnetic field and heat conduction under more realistic conditions in the future.

\par
\par
\par
\par
\par
\par

$\\$
{\bf Availability of Data}
$\\$
The data that support the findings of this study are available from the corresponding author upon reasonable request.

\bibliographystyle{apsrev}

\end{document}